# Entanglement and Quantum Nonlocality Demystified.

Marian Kupczynski

*Département de l'Informatique, UQO, Case postale 1250 ,succursale Hull, Gatineau. Quebec, Canada J8X 3X 7*

**Abstract.** *Quantum nonlocality* is presented often as the most remarkable and inexplicable phenomenon known to modern science. It has been known already for a long time that the probabilistic models used to prove Bell and Clauser-Horn-Shimony-Holt inequalities (BI-CHSH) for spin polarization correlation experiments (SPCE) are incompatible with the experimental protocols of SPCE. In particular these models use the same common probability space, joint probability distributions and/or conditional independence to describe coincidence experiments in incompatible experimental settings. Strangely enough these results are not known or simply neglected. This is why we will once again reanalyze Bell locality assumptions and show that they have nothing to do with the notion of *Einsteinian locality* therefore their violation should not be called *quantum nonlocality* but rather *quantum non-Kolmogorovness* or *quantum contextuality*. Moreover, if local variables describing measuring instruments are correctly taken into account then BI-CHSH can no longer be proven and one can easily construct non-signaling probabilistic models able to reproduce the predictions of QT. The violation of BI-CHSH is considered usually as a proof that a quantum state is entangled. Since BI-CHSH are violated also in some experiments from outside the domain of quantum physics therefore the entanglement is not exclusively a quantum phenomenon. In view of these arguments the further testing of BI-CHSH inequalities in search for the loopholes does not seem to be necessary.

## INTRODUCTION

**This is an update of the article published in 2012. In particular we explain shortly that in the experiments discussed in the last section of this paper the inequality which is violated it is not Bell inequality. We add Eq.12 and the reference to the recent paper [66] in which a detailed and correct discussion of these experiments is given. We correct also an insignificant misprint in Eq.7 and Eq.8. All other arguments remain unchanged.**

The entanglement and the quantum nonlocality are considered to be the greatest mysteries of the Nature [1-3, 5]. In this paper we want to give convincing arguments that there is nothing mysterious in the fact that BI-CHSH [4-7] are violated in spin polarization correlation experiments (SPCE) [8-10].

The true meaning and limitations of so called Bell's ``locality`` assumptions were pointed out and discussed by several authors [11-46]. The various models able to reproduce the predictions of quantum theory (QT) for SPCE were constructed [12, 14, 31, 36, 40, 41, 47, 48, 64]. Experiments violating BI from various domains of science were found [26, 27, 46, 48, 64].

It was even possible to simulate with success in a local and consistent way several experiments in quantum optics including SPCE and in neutron interferometry [47, 62].

Strangely enough these results are neglected by the physical community. We hope that this paper will finally make a difference.

We will show that the correlations between the outcomes of the measurements performed on two non- interacting quantum systems can be explained by partially preserved memory of their common "history" carried by some local variables.

Finally we will give an example of realizable macroscopic coincidence experiments performed on an entangled pair of "classical qubits'' which violate BI. **This violation is only apparent please see Eq.12 and [66].**

## LONG RANGE CORRELATIONS IN EPR-TYPE EXPERIMENTS.

One of the assumptions in the orthodox QT was the *irreducible randomness* of the acts of the measurements which seemed to imply that the results of faraway local measurements performed at the same moment of the laboratory time on two non-interacting physical systems should be uncorrelated.

In the famous EPR paper the authors [49] showed that according to QT two quantum systems which interacted in the past and after evolved freely were in a state which was a non-convex sum of different tensor product states. In such a state, called nowadays entangled, the results of the local experiments performed on the members of the EPR pair were strongly correlated contradicting the arguments given above [34].

Since QT gave only the prediction for the existence of these correlations without providing any explanation how they could be produced. Bell said: *"The correlations cry for explanation"* and he was right. In classical physics it is well known that the long-range correlations between the measurements performed on two non-interacting physical systems can be correlated for various reasons and one of them is their common history and various conservation laws.

Therefore one way to explain the quantum long range correlations was to assume that there existed some supplementary variables describing the behavior of the individual physical systems during the experiment in which the memory of their common history could be stored. The interpretation of QT which allows for the existence of such variables is a statistical interpretation [11-39, 50, 51] of QT proposed and strongly advocated by Einstein.

In the modern exposition of the statistical interpretation of QT [22,26,27,32-35,50,51] the wave function $\Psi$ or a density operator $\rho$ do not describe a single physical system but an ensemble of equivalent preparations of physical system and if the observable O is measured both the preparation and the experimental settings are the contexts under which the conditional probabilities (propensities) $P(a|(\Psi, \hat{O})$ or $P(a|\rho, \hat{O})$ , where $\hat{O}$ is an operator representing the observable O or an appropriate POVM, are calculated and compared with the experimental data. Therefore probabilities are neither attributes of individual physical systems nor attributes of measuring devices but only characteristics of a whole random experiment. The QT is not a theory of quantum individual systems but it is a contextual theory of quantum phenomena.

In SPCE which are EPR-type experiments one is assuming that a source S is producing "pairs of photons" in a singlet state, the "photons" separate and hit the faraway polarization filters and detectors. We will continue to use this *mental picture* even if nobody knows what *mental picture* of a photon is appropriate [51].

Fortunately in order "to understand" long range correlations in SPCE we do not need a detailed description of the time evolution of supplementary variables but we need plausible probabilistic models able to reproduce QT predictions.

Bell [4, 5] proposed two probabilistic hidden variable models of SPCE and found that they led to BI-CHSH violated by some predictions of QT. Bell did not notice that his probabilistic models were inconsistent with the experimental protocols of SPCE.

The intimate relation between experimental protocol and a probabilistic model used to describe it was discovered many years ago by Bertrand.

## BERTRAND'S PARADOX.

In 1889 Bertrand [52, 53] discovered the following paradox. If we consider two concentric circles on a plane with radii R and R/2, respectively we can ask a question:" What is the probability P that a chord of the bigger circle chosen at random cuts the smaller one in at least one point?"

The various answers seem to be equally reasonable. If we divide the ensemble of all chords into sub-ensembles of parallel chords, we find P=1/2. If we consider the sub-ensembles of chords having the same beginning, we find P=1/3. Finally if we look for the midpoints of the chords lying inside the small circle, we find P=1/4 .The solution of the paradox is simple.

Different probabilistic models leading to different answers correspond to different protocols of the random experiments performed to find the answer to the Bertrand's question. It proves the contextual character of the probabilities and their intimate relation to specific random experiments [26, 27, 30-35]. Therefore the probability of obtaining 'head' in a coin flipping experiment using a particular flipping device is neither a property of a coin nor a property of a flipping device. It is only the characteristic of the whole random experiment: "flipping this particular coin with that particular flipping device". Similarly the quantum state vector should not be treated as an attribute of an individual physical system.

## BELL'S LOCALITY ASSUMPTION - "STOCHASTIC" LSHV.

Let us assume following [2] that in SPCE a source S is emitting the correlated pairs of photons which after the passage through the experimental set-up are hitting the polarization measuring instruments x and y. The binary outputs *a* and *b* on the detectors can be interpreted as the values of corresponding random variables A and B. The hidden variables of the probabilistic model called in [2] local variables are λ and since the outputs are correlated:

$$P(\mathrm{a,b}|\mathrm{x,y}) = P(\mathrm{A=a,\ B=b}\,|\,\mathrm{S,x,y}) \neq P(\mathrm{A=a}\,|\,\mathrm{S,x,y})\,P(\mathrm{B=b}\,|\,\mathrm{S,x,y}) \tag{1}$$

Let us now analyze the assumptions of the probabilistic model leading to BI-CHSH as presented in [2]:

$$P(a,b\,|\,x,y)=\sum_{\lambda\in\Lambda}P(\lambda)P(a\,|\,x,\lambda)P(b\,|\,y,\lambda) \qquad (2)$$

The probabilities $P(a,b\,|\,x,y)$ are calculated as weighted averages of the probabilities $P(a,b\,|\,x,y,\lambda)$ from the experiments labeled by λ and are assumed to factorize:

$$P(a,b\,|\,x,y,\lambda)=P(a\,|\,x,\lambda)P(b\,|\,y,\lambda) \qquad (3)$$

**The true meaning of Eq.3 which is called *Bell's locality* condition is the *assumption of conditional independence* of the random variables A and B describing the outcomes of the experiments x and y for a fixed value of λ.**

Since it is well known that CHSH inequalities are violated [8-10] and the experimental settings can be freely and randomly chosen by experimenters the authors concluded that *Bell's nonlocality* is an established experimental fact.

We agree that the violation of Eq.3 is a an established experimental fact but *Bell's nonlocality* or *quantum nonlocality* should be called *quantum non-Kolmogorovness* as suggested by Accardi [11-14] and Khrennikov [22-27] or *quantum contextuality* [34-35]. Besides the probabilistic model presented above is inconsistent with experimental protocol used in SPCE.

Namely since Λ and P (λ) in Eq.2 do not depend on x and y thus the local variables λ describe only pairs of photons in the moment when they hit polarization measuring instruments x and y. In each pair of experiments (x, y) we have the same statistical ensemble of pairs of photons.

Following Bertrand we see that the Eq.2 and Eq.3 specify a particular experimental protocol which could be used in order to find the correlations between the observed outcomes $a=\pm1$ and $b=\pm1$ for a chosen pair:

1. Repeat the experiments x on a first photon from a given pair several times and find an estimate $\hat{P}(a\,|\,x,\lambda)$ of the probability $P(a\,|\,x,\lambda)$. From the repeated experiments y on the second photon find the estimate $\hat{P}(b\,|\,y,\lambda)$.
2. Find the estimate: $\hat{P}(a,b\,|\,x,y,\lambda)=\hat{P}(a\,|\,x,\lambda)\;\hat{P}(b\,|\,y,\lambda)$.
3. Repeat the steps 1 and 2 for the next pair of photons
4. Average the results obtained after many repetitions of the preceding steps and obtain $\hat{P}(a,b\,|\,x,y)$ being an estimate of $P(a,b\,|\,x,y)$.

This protocol has nothing to do with the protocol of the coincidence experiments in SPCE and even in theory it could not be implemented since we cannot keep a pair of photons and repeat the experiments on each of its members. Besides in such a protocol any correlations due to a common history would be lost.

To be able to describe SPCE we have to modify the probabilistic model from Eq. 2:

$$P(a,b\,|\,x,y)=\sum_{\lambda\in\Lambda_{xy}}P(\lambda)P(a\,|\,x,\lambda_1,\lambda_x)P(b\,|\,y,\lambda_2,\lambda_y) \qquad (4)$$

**In this local probabilistic model the probability space $\Lambda_{xy}$ is different for each pair of experiments** because it contains not only the local variables describing the pairs but also the local variables describing the instruments $\lambda=(\lambda_1, \lambda_2, \lambda_x, \lambda_y)$ . Nobody should be tempted to think that the hidden variables describing polarization filters, as perceived by the signals, should not change when we rotate them . For a laser pulse the rotated crystal lattice looks completely different when rotated.

In order to preserve the memory of the common history the values *a* and *b* have to be strictly determined for each choice of the values of the parameters λ thus: $P(a,b \mid x,y,\lambda) = P(a \mid x,\lambda_1,\lambda_x)P(b \mid y,\lambda_2,\lambda_y)$ are 0 or 1 and the experimental protocol is consistent with SPCE in contrast to the protocol with non-vanishing $P(a,b \mid x,y,\lambda)$.

The model is explicitly non-signaling, all the correlations are coded in P(λ) and since in Eq.4 we have additional specific local parameters for each (x, y) one can reproduce the correlations predicted by QT and violating BI-CHSH. Please note that the *local choices of the experimental settings made by the experimenters are completely independent* and the order in which they are made has no influence on the correlations observed.

We used above discrete local variables but the similar reasoning applies for continuous local variables what we will do below discussing the first proof of BI given by Bell in the language of correlation functions [4].

## BELL'S LOCALITY ASSUMPTION - "REALISTIC" LRHV.

The random experiments (x, y) can be labeled by their orientation vectors (**A**, **B**). The outputs of the experiments (**A**, **B**) are the values ±1 of the corresponding binary random variables A and B such that the expectation values E(A)=E(B)=0. The correlation between the outcomes are measured by the covariance function Cov(A,B)= E(AB)-E(A)E(B) which in our case reduces to the correlation function E(AB). Bell [4] introduced local realistic hidden variable (LRHV) probabilistic model in which photon pairs are described by $\lambda \in \Lambda$ and the random variables are represented by bi-valued functions A(λ)=±1 and B(λ)= ±1 and

$$E(AB) = \int_{\Lambda} A(\lambda)B(\lambda)\rho(\lambda)d\lambda . \qquad (5)$$

It was emphasized already by Kolmogorov that any given random experiment has its own probability space. Therefore any correlations in the experiment (x, y) can be reproduced without violating locality by using a corrected equation:

$$E(AB) = \int_{\Lambda_{xy}} A(\lambda)B(\lambda)\rho_{xy}(\lambda)d\lambda . \qquad (6)$$

**In Eq.6 the probability space $\Lambda_{xy}$ and the probability density $\rho_{xy}$ are different for each couple of the experiments (x, y).** Bell claimed that the replacements $\Lambda_{xy} = \Lambda$ and $\rho_{xy}(\lambda) = \rho(\lambda)$ are justified by the locality of experiments (x, y) .

Bell and his followers were wrong because **Eq. 5 means only that we may use the unique probability space and the joint probability distribution to find the marginal probability distributions for all experiments (x, y) performed in** **incompatible experimental settings.** In particular according to Eq. 5 we can find correlation functions for any n-tuple of random variables which cannot be performed in SPCE. For example if we study the correlations of outcomes in four pairs of experiments ($\mathbf{A}_1$, $\mathbf{B}_1$), ($\mathbf{A}_1$, $\mathbf{B}_2$), ($\mathbf{A}_2$, $\mathbf{B}_1$) and ($\mathbf{A}_2$, $\mathbf{B}_2$) and apply Eq.5 we may also find $E(A_1A_2B_1B_2)$ which is meaningless because we cannot measure at the same time the spin polarization projections in two different directions on any single photon and we do not have the joint probability distribution which is needed.

### Common Probability Space and Joint Probability Distributions.

Several mathematicians and physicists noticed that one should not use the common probability space and joint probability distributions to describe SPCE. This is the reason why BI-CHSH are violated not the non- locality of QT. Let us mention here few of them in the alphabetic order: Accardi [11-14], Fine [18, 19], Hess and Philip [20,21], Khrennikov [22-27], Kupczynski [28-35], Matzkin [37],de Muynck, de Baere and Martens [17], Newenhuizen [38,39], Pitovsky [41,42], Rastal [43], de Raedt, Hess, and Michielson [46]. Due to the increased interest in the subject the important results of Boole and Vorob'ev were found.

In 1862, Boole [54] showed that whatever process generates a data set *S* of triples of variables ($S_1$, S2, S3) where $S_i = \pm 1$, then the averages of products of pairs $S_iS_j$ in a data set S have to satisfy the equalities very similar to BI [24, 27, 42, 46].To prove Boole's inequalities it is essential that all pairs are selected from one and the same set of triples. If we select pairs from three different sets of pairs of dichotomous variables, then Boole's inequalities cannot be derived. If we have a data set of triples we can estimate the joint probability distribution for the triples and the corresponding marginal probability distributions. Therefore we see that **BI may be interpreted as a necessary condition for the existence of the joint probability distributions of the values of dichotomous random variables which can be measured pair-wise but not simultaneously.**

Similar question was studied in 1962, before BI were known, by Vorob'ev [55] who asked a question: *"Is it possible to construct the joint probability distribution for any triple of pair-wise measurable dichotomous random variables?"*. The answer was *"No* "and he gave simple counterexamples [24, 26, 27, 55]. In general the joint probability distributions of random variables do not exist if we cannot measure them simultaneously or in the sequence without affecting the preceding outcomes. Neither Boole nor Vorob'ev mentioned the *realism* and/or the *locality*. Therefore the violation of BI does not allow us to question the locality of physical interactions.

### Local Realism and Contextuality.

The violation of BI-CHSH allows us to reject a notion of LRHV as introduced by Bell. By writing Eq. 5 Bell assumed a particular description of the measurement process in SPCE. He assumed that *the spin polarization projections in any direction for each photon, unknown to the observer, are predetermined at the source S and the polarization analyzing instruments x or y "read" only their preexisting values*. One can even incorporate the reading errors allowing the random variables to take also a value zero for some values of $\lambda$.

This definition of the *realism* comes from classical physics. For example let us consider a box which contains in different proportion balls which can be small or big, blue or white, made from metal or from wood. We can draw a ball from a box, record its properties and replace it. Since each ball is characterized by three classical attributes then there exists a joint probability distribution of three random variables (color, size, and material) and Eq.5 may be used to find any binary correlation functions. The spin polarization projections in all directions are not simultaneously measurable attributes of the photon therefore the probabilistic model defined by Eq.5 cannot describe correctly SPCE. The quantum systems do not resemble painted balls but rather Accardi's *chameleons* [12-14].

It has been known from the early days of quantum physics that the *realism understood as above* was inconsistent with QT and led to various paradoxes. The hidden variables able to reproduce the predictions of QT have to be *contextual* (depending on the experimental context) what was proven using different approach by Gleason [56], Kochen Specker [57], Geenberger et al. [58] and Mermin [59] . The recent review of a vast research stimulated by these papers was published by Liang et al. [60]. It could be noticed already in 1970 when Wigner [61] used realistic hidden variables to study the passage of photons through a sequence of three incompatible polarization filters and proved Bell- type inequalities which were violated by some predictions of QT. The locality assumption was completely irrelevant in his proof.

The impossibility of local realistic hidden variables used by Bell does not mean that the physical systems and the measuring devices are not real [3,39]. It means only that the measurement process is not a passive reading of the preexisting values of the physical observables. Therefore *any attempt to deduce quantum statistical predictions using additional variables must be contextual in the sense that it has to incorporate the additional variables of the measuring devices* like it was done for example in Eq.4 and Eq.6. Please note that in spite of the fact that the author of [39] arrived to the same conclusions concerning the violation of BI-CHSH he was using the term *contextuality* in exactly opposite meaning. Unfortunately the *realism* and the *contextuality* have often different meaning for different authors what leads to confusion.

## LONG RANGE CORRELATIONS EXPLAINED

By introducing the contextual hidden variables carrying the memory of the common past one can in an intuitive way explain the existence of long range correlations in SPCE.

Namely the signals leaving a source S are described by a random variable C. After passing by the experimental set-up when they arrive to the measuring instruments (x, y) in distant locations they are described by the random variables $C_1(C)$ and $C_2(C)$ respectively. The measuring devices, *as perceived by the signals*, are described by the random variables $X_1$ and $X_2$. The clicks on the detectors in the experiment (x, y) are described by the random variables A and B.

*The correct non-signaling locality assumption implies* $A=A(C_1(C), X_1)$ *and* $B=B(C_2(C), X_2)$ thus the random variables A and B being the functions of the same random variable C describing the signals as they were sent by the source S are not independent but correlated.

The functions $A(C, X_1)$ and $B(C, X_2)$ are complicated functions of C but the long range correlations also exist if A and B are much simpler functions of C.

For example if A=sC and B=tC where s and t are positive real numbers and C is a random variable such that; $E(C)=\int_{-\infty}^{\infty} x f_C(x)dx$ then the probability densities of the variables A and B are simple functions of the probability density of C:

$$f_A(x_1)=\frac{1}{s}f_C(\frac{x_1}{s}) \quad and \quad f_B(x_2)=\frac{1}{t}f_C(\frac{x_2}{t}) \tag{7}$$

It is easy to find the joint probability density $f_{AB}(x_1,x_2)$ of A and B:

$$f_{AB}(x_1,x_2)=\frac{1}{s}f_C(\frac{x_1+x_2}{s+t})\delta(x_2-\frac{t}{s}x_1). \tag{8}$$

**In this update we corrected misprints in Eq.7 and Eq.8.** Namely in Eq.7 we changed $f_A(\frac{x_2}{t})$ into $f_C(\frac{x_2}{t})$ and in Eq.8 we changed $f_A(\frac{x_1+x_2}{s+t})$ into $f_C(\frac{x_1+x_2}{s+t})$. Using this probability density we can find the correct marginal probability densities from Eq.7. The A and B are strongly correlated because $f_{AB}(x_1,x_2)\neq f_A(x_1)f_B(x_2)$ and $E(AB)\neq E(A)E(B)$.

## CLASSICAL "TWO QUBIT" SYSTEM.

Let us consider now a simple EPR type experiment realizable in the laboratory with macroscopic physical systems which can be described in a similar way as above.

We choose the following protocol for this experiment.

1. We place metal balls having the same dimensions but different masses M and m, with M>m, in some fixed places $P_1$ and $P_2$ on the horizontal perfectly smooth surface.
2. A device D, with built in random numbers generator, is imparting a constant rectilinear velocity with a speed described by a random variable V distributed according to a probability density $f_V(x)$ and the ball is sliding without friction and without rotating towards the heavier ball.
3. After elastic head on collision the balls rebound with the speeds $V_1$ and $V_2$ for heavier and lighter ball respectively and are sliding without friction on the same straight line but in the opposite directions. Devices $D_1$ and $D_2$ measure locally these speeds and record them and we go back to the step1.

4. The random variables are strongly correlated even if the collision is not perfectly elastic and for the perfect elastic collision we get:

$$V_1 = \frac{2m}{m+M} V \ and \quad V_2 = \frac{M-m}{m+M} V \tag{9}$$

We can also connect to the devices $D_1$ and $D_2$ detectors A and B respectively which assign the binary outcomes $\pm 1$ in a deterministic way in function of the observed values of $V_1$ and $V_2$. We may interpret the values of these outcomes as the values of random variables A and B which are also strongly correlated.

Please note that if the head on collision is not perfectly elastic Eq.9 does not hold and since we don't know a value of V before the collision and how imperfectly elastic the collision was the values $V_1$ and $V_2$ are not predetermined at the step 1 but they are only known after the final measurements are done. If we do not know how the devices $D_1$+A and $D_2$ +B function and we only observe the clicks then V,$V_1$ and $V_2$ are our hidden variables helping us to "understand" the long range correlations of faraway clicks.

Let us now consider a particular experiment performed according to the above protocol in which a pair of the masses will behave as a "classical entangled two qubit system".

We assume that M=4m and that the collision is perfectly elastic and that $f_V(x)$ is a symmetric probability density on an open interval ]0, 10[.

From Eq.9 we obtain $V_1 = \frac{2}{5}V$, $V_2 = \frac{3}{5}V$ and $V_1 = \frac{2}{3}V_2$ allowing us to find probability densities $f_{V_1}(x)$ and $f_{V_2}(x)$ defined on the intervals] 0, 4[ and ] 0, 6[ respectively.

Finally we define the functioning of three detectors A, B and C which can be attached to the devices $D_1$ and $D_2$:

- $A(x) = -1$ if $0 < x < 2$ and $A(x) = 1$ if $2 \leq x$
- $B(x) = -1$ if $0 < x < 3$ and $B(x) = 1$ if $3 \leq x$
- $C(x) = 1$ if $0 < x < 3$ and $C(x) = -1$ if $3 \leq x$

We see that:

- if $V_1 < 2$ then $V_2 < 3$ and $A(V_1)B(V_2) = (-1)(-1) = 1$
- if $2 \leq V_1$ then $3 \leq V_2$ and $A(V_1)B(V_2) = (1)(1) = 1$.

Thus E(AB)=1 and from the definition of B and C we get immediately E(AC)=-1.

Now we may find E (BC) what is slightly more difficult:

- if $V_1 < 2$ then $V_2 < 3$ and $B(V_1)C(V_2) = (-1)(1) = -1$
- if $2 \leq V_1 < 3$ then $3 \leq V_2 < 4.5$ and $B(V_1)C(V_2) = (-1)(-1) = 1$
- if $3 \leq V_1$ then $4.5 \leq V_2$ and $B(V_1)C(V_2) = (1)(-1) = -1$

therefore

$$E(BC) = \int_2^3 f_{V_1}(x)dx - \int_0^2 f_{V_1}(x)dx - \int_3^4 f_{V_1}(x)dx \tag{10}$$

From Eq.10 and using the symmetry and the positivity of the probability density function on ]0,4[ we may write $E(BC) = -1 + h$ where $0 < h < 1$.

We see immediately that one of BI [30]:

$$| E(AB) - E(AC) | \leq 1 + E(BC) \qquad (11)$$

is strongly violated in our experiments because $|1-(-1)|> h$ where $0<h<1$. If on the right hand side of Eq.11 we put 1-E(BC) we get $2>2-h$ and another inequality violated.

The *violation of BI is considered as the most reliable proof of the entanglement* [63] so we have to conclude that we prepared our two classical masses in an entangled state.

**This conclusion is incorrect because the inequality (11) looks as the Bell inequality but in fact it contains 4 different random variables instead of 3.**

$$| E(A(\mathrm{V}_1)B(V_2) - E(A(\mathrm{V}_1)C(V_2) | \leq 1 + E(B(\mathrm{V}_1)C(\mathrm{V}_2)) \qquad (12)$$

**A detailed discussion of these experiments and of the inequalities (11) and (12) may be found in (66).**

## CONCLUSIONS

Instead of underlying the mysterious character of the long range correlations in SPCE as coming from out of space-time we gave a rational explanation of these correlations by a *partial memory* of their common history kept and carried by the "individual physical systems". In order to preserve this *partial memory* the apparently randomly produced outcomes of quantum measurements must be in fact obtained in some uncontrollable but deterministic process depending on some additional variables carrying the *"memory"* and on some variables, which we call *contextual*, describing internal states of the measuring apparatus during the measurement. The probabilistic models using such variables can reproduce all the probabilistic predictions of QT and do not allow proving BI-CHSH inequalities.

Even if such models are not needed, unless some new experimental predictions could be obtained using them, their existence and the existence of the computer simulations [47, 62] allow us to demystify quantum phenomena by providing some "intuitive understanding".

BI-CHSH cannot be proven due to the non-existence of the common probability space and joint probability distributions for the coincidence experiments in incompatible experimental settings. Strangely enough the non-existence of the joint probability distributions for noisy spin measurements was recently proven in [60] without giving any reference to [11-46]. Since the *quantum nonlocality* has nothing to do with the non-locality of Nature it should be called differently.

BI-CHSH are violated also in the experiments from outside the domain of quantum physics [48, 26, 27, 63, 64] therefore the entanglement is not exclusively a quantum phenomenon.

Of course we do not question the usefulness of the long range correlations characterizing the entangled physical systems in the domain of Quantum Information.

However one should not forget that the anti-correlations are never strict [31, 35], that the wave function should not be treated as an attribute of the individual quantum system which can be changed instantaneously and that the unperformed experiments have no results.

Hoping that in future the physical community will take seriously the ideas reviewed above we finish this article with the words of Jaynes [62] : ”He who confuses reality with his knowledge of reality generates needless artificial mysteries.”

## ACKNOWLEDGMENTS

I would like to thank Andrei Khrennikov for the invitation and the warm hospitality extended to me during this interesting conference. I would like also to thank UQO for a travel grant.